\documentstyle{article}
\def\beq{\begin{equation}}
\def\beqa{\begin{eqnarray}}
\def\eeq{\end{equation}}
\def\eeqa{\end{eqnarray}}
\newcommand{\text}[1]{\mbox{\scriptsize{#1}}}

\title{Monte Carlo simulation of a hard-sphere gas in the planar Fourier 
flow with a gravity field}
\author{E. E. Tahiri\thanks{D\'epartement de Math\'ematique, 
Universit\'e Moulay Isma\"{\i}l,
Mekn\`es, Morocco; e-mail: tahiri@fsmek.ac.ma}, M. Tij\thanks{D\'epartement 
de Physique, Universit\'e Moulay Isma\"{\i}l,
Mekn\`es, Morocco; e-mail: mtij@fsmek.ac.ma} and A. 
Santos\thanks{Departamento de F\'{\i}sica, Universidad de Extremadura, 
E-06071 Badajoz, Spain; e-mail: andres@unex.es}}\date{\today}
\begin{document}
\maketitle
\begin{abstract}
By means of the Direct Simulation Monte Carlo method, the Boltzmann equation 
is numerically solved for a gas of hard spheres enclosed between two 
parallel plates kept at different temperatures and subject to the action of 
a gravity field normal to the plates.
The profiles of pressure, density, temperature and heat flux are seen to be 
quite sensitive to the value of the gravity acceleration $g$. If the gravity 
field and the heat flux are parallel ($g>0$), the magnitudes of both the 
temperature gradient and the heat flux are smaller than in the opposite case 
($g<0$). When considering the actual heat flux relative to the value 
predicted by the Fourier law, it is seen that, if $g>0$, the ratio increases 
as the reduced local field strength increases, while the opposite happens if 
$g<0$. The simulation results are compared with theoretical predictions for 
Maxwell molecules.
\end{abstract}

\section{Introduction}
\label{sec:1}
The steady planar Fourier flow  is one of the basic nonequilibrium states. 
It corresponds to a macroscopic system enclosed between two parallel 
infinite plates located at $z=0$ and $z=L$ and kept at different 
temperatures $T_-$ and $T_+$. After a certain transient period, the system 
reaches a steady state characterized by a temperature gradient 
$\partial T/\partial z$ along the direction $z$ normal to the plates and a 
constant heat flux $q_z$. In the case of a fluid described by the 
Navier-Stokes equations, the heat flux is just proportional to the 
temperature gradient (Fourier law):
\beq
\label{1.1}
q_z^{\text{NS}}=-\kappa(T(z))\frac{\partial T}{\partial z},
\eeq
where $\kappa(T)$ is the thermal conductivity coefficient, which, in 
general, depends on the local temperature $T$.
In principle, the validity of the linear relation (\ref{1.1}) is 
restricted to small gradients, i.e. to $\ell\gg \lambda$, where $\ell$ is 
the characteristic {\em hydrodynamic\/} length defined as $\ell=T|\partial 
T/\partial z|^{-1}$ and $\lambda$ is a {\em microscopic\/} 
characteristic length (such as the mean free path in the case of a dilute 
gas). However, computer simulations of both dense fluids \cite{CT80} and 
dilute gases \cite{MKBCN87,MASG94}, as well as  kinetic theory treatments 
\cite{AMN79,SBG86,SBKD89,KDSB89,KD89} show that equation (\ref{1.1}) is an 
excellent approximation even if $\ell\sim \lambda$. In the special case of 
dilute gases, Asmolov {\em et al.\/} \cite{AMN79} found an exact solution of 
the Boltzmann equation for Maxwell molecules \cite{note} in which equation 
(\ref{1.1}) is verified for {\em arbitrary\/} values of the thermal 
gradient. The same result is obtained from an exact solution \cite{SBG86} of 
the Bhatnagar-Gross-Krook (BGK) kinetic model \cite{BGK54} for any 
interaction potential.

The effect of gravity on the heat conduction of {\em dilute}
gases is usually neglected. This is because the characteristic distance 
associated with gravity (i.e. the scale height $h=v_0^2/g$, where $v_0$ is 
the thermal velocity and $g$ is the gravity acceleration) is  much larger 
than  $\ell$ and $\lambda$ under usual laboratory conditions. Thus, the 
Fourier law (\ref{1.1}) still holds if
$h\gg \ell\sim\lambda$.	 
On the other hand, an interesting question from a physical point of view is 
whether or not the heat conduction is influenced by
 a gravity field ${\bf g}$ along the $z$-direction when the 
conditions of rarefaction and/or the strength of the  field are such
that the ratio $\lambda/h$ is not negligibly small. 
In the case of  Earth's atmosphere, for instance, $h$ varies only within 
the range of 5--10~km up to an altitude of 100~km \cite{IC80,C78}. On the 
other hand, $\lambda$ increases from $10^{-5}$~cm at the surface to tens of 
km at an altitude of 500~km. Consequently, $\lambda/h\sim 10^{-11}$ at the 
surface but rapidly increases with the altitude, being $\lambda/h\sim 1$ at 
the base of the exosphere \cite{IC80}.

In the study of the gravity effect on the heat conduction the main quantity 
of interest is the {\em reduced thermal conductivity\/} coefficient 
$\kappa^*(\epsilon,g^*)$, where
\beq
\label{1.2}
\kappa^*\equiv \frac{q_z}{q_z^{\text{NS}}},
\eeq
\beq
\label{1.3}
\epsilon\equiv \frac{\lambda}{\ell}
=\lambda\left|\frac{\partial\ln T}{\partial z}\right|,
\eeq
\beq
\label{1.4}
g^*\equiv \frac{\lambda}{h}
=\frac{\lambda g}{2k_BT/m}.
\eeq
The quantity $\epsilon$ is a dimensionless measure of the  
thermal gradient over the scale of the mean free path, while  $g^*$ is a measure
of the strength of the gravity field over the same scale. In equation 
(\ref{1.4})
we have identified the thermal velocity as $v_0=(2k_BT/m)^{1/2}$, where $k_B$
is the Boltzmann constant and $m$ is the mass of a particle. 
As a convenient definition of the mean free path we take
\beq
\label{1.5}
\lambda=\frac{2 m\kappa v_0}{5k_B p},
\eeq
where $p=nk_BT$ is the hydrostatic pressure, $n$ being the number density.
The above definition of $\lambda$ is based on the result for the thermal 
conductivity coefficient $\kappa$ in the BGK model \cite{CC70,Cer90}.
It must be pointed out that all the quantities in (\ref{1.2}--\ref{1.4}) are
{\em local\/}, i.e. they depend on $z$. For instance, $\kappa^*$ is the ratio
between the actual value of the heat flux (which is in fact uniform in the 
steady state) and the local value obtained from (\ref{1.1}) with the actual
local values of the temperature and its gradient. If the system is large enough
($L\to\infty$) and we restrict ourselves to the {\em bulk domain\/} (i.e. 
far from the boundaries),
it is expected that all the space dependence of $\kappa^*$ occurs through 
$\epsilon$ and $g^*$ only.

The problem of elucidating the effect of $g^*\neq 0$ on the effective thermal
conductivity $\kappa^*$ has been
recently addressed by us from a theoretical point of view 
\cite{TGS97,TGS99b,DST99,TGS99}. In Ref.\ \cite{TGS97}  the 
Boltzmann equation for Maxwell molecules was solved by a perturbation 
expansion through order $g^2$. The result is
\beq
\label{1.6}
\kappa^*(\epsilon,g^*)=	 \kappa^*(\epsilon,0)+\frac{46}{5}\epsilon g^*+
\left(\frac{24}{5}+503.7\epsilon^2\right){g^*}^2+{\cal O}({g^*}^3),
\eeq
where $\kappa^*(\epsilon,0)=1$.
Henceforth we take the convention that $g>0$ when the field ${\bf g}$ is 
parallel to
the heat flux vector ${\bf q}$ and $g<0$ when ${\bf g}$ is antiparallel to ${\bf q}$.
In the first case, according to equation (\ref{1.6}), the heat transport is 
enhanced with respect to the Navier-Stokes prediction ($\kappa^*>1$), while 
it is inhibited
if $g<0$. Of course, in the limit of a negligible field ($g^*\to 0$), the 
validity of the Fourier 
law for arbitrary $\epsilon$ \cite{AMN79} is recovered ($\kappa^*\to 1$).
A similar analysis has been carried out in the context of the BGK model of
the Boltzmann equation, also for Maxwell molecules \cite{TGS99b}. 
This allowed us to perform the expansion through sixth order in the field, 
the behaviour of the numerical coefficients indicating that the series 
expansion is only {\em asymptotic}. The result to second order is
\beq
\label{1.7}
\kappa^*(\epsilon,g^*)=	\kappa^*(\epsilon,0)+\frac{58}{5}\epsilon g^*+
\left(\frac{32}{5}+\frac{47\,968}{25}\epsilon^2\right){g^*}^2+{\cal 
O}({g^*}^3).
\eeq
The asymptotic analysis of Ref.\ \cite{TGS99b} agrees well with a 
finite-difference numerical solution of the BGK equation \cite{DST99}.
Comparison between equations (\ref{1.6}) and (\ref{1.7}) shows that the BGK 
model
tends to overestimate the influence of the gravity field.
A much more complex gravity effect appears in the planar Couette flow 
\cite{TGS99}, where normal and shear stresses are present in addition to heat
transport.

Thus far, all the previous studies about
the  influence of gravity on the heat 
conduction have been restricted to  {\em Maxwell molecules}. In addition, 
most of them have been based on theoretical asymptotic analyses of the 
Boltzmann or BGK equations, the investigation of Ref.\ \cite{DST99} being 
the only exception.
The main merit of the Maxwell interaction is that it usually makes 
the analytic treatment of the Boltzmann equation more manageable, but 
otherwise it is a rather unrealistic potential. Structural, 
thermodynamic and transport properties of real fluids are much better 
captured by the hard-sphere model \cite{HM86}. 
The aim of this paper is to investigate the gravity effect on the 
planar Fourier flow in the case of a dilute gas of {\em hard spheres\/} by 
solving numerically the Boltzmann equation by means of the DSMC method 
\cite{Bird}. 
As will be seen, the simulation results agree qualitatively with the 
theoretical analyses.
The physical problem is stated in section \ref{sec:2}, where a 
special attention is paid to the choice of the boundary conditions. The 
simulation method is  described in section \ref{sec:3}. Section 
\ref{sec:4} presents  the results. Finally, the conclusions are 
summarized in section \ref{sec:5}.

\section{Planar Fourier flow in the presence of a gravity field}
\label{sec:2}
Let us consider a dilute gas of hard spheres enclosed between two parallel plates located at
$z=0$ and $z=L$. Both plates are maintained at temperatures $T_-$ and $T_+$,
respectively.
Without loss of generality, we will assume that $T_+> T_-$. In addition, a
constant gravity field ${\bf g}=-g\widehat{\bf z}$ is applied.
Figure \ref{fig0} presents a schematic illustration of the system geometry.
Under these conditions, the Boltzmann equation reads \cite{CC70,Cer90}
\beqa
\label{2.1}
\left(\frac{\partial}{\partial t}
+v_z\frac{\partial}{\partial z}
-g\frac{\partial}{\partial v_z}\right)
f(z,{\bf v},t)&=&\sigma^2\int d{\bf v}_1\int 
d\widehat{\mbox{$\boldmath{\sigma}$}} \,\Theta({\bf g}\cdot
\widehat{\mbox{$\boldmath{\sigma}$}})({\bf 
g}\cdot\widehat{\mbox{$\boldmath{\sigma}$}}) 	  \nonumber\\
&&\times \left[	f(z,{\bf v}',t)f(z,{\bf v}_1',t)\right.
\nonumber\\
&&\left.-
f(z,{\bf v},t)f(z,{\bf v}_1,t)\right].
\eeqa
Here, $f(z,{\bf v},t)$ is the one-particle velocity distribution function,
$\sigma$ is the diameter of a sphere,
$\Theta$ is the Heaviside step  function, ${\bf g}={\bf v}-{\bf v}_1$ 
is the relative velocity, $\widehat{\mbox{$\boldmath{\sigma}$}}$ is a unit 
vector in the direction of the line joining the centers of the two colliding 
particles, and ${\bf v}'={\bf v}-
({\bf 
g}\cdot\widehat{\mbox{$\boldmath{\sigma}$}})\widehat{\mbox{$\boldmath{\sigma}$}}$ 
and ${\bf v}_1'={\bf v}_1+({\bf 
g}\cdot\widehat{\mbox{$\boldmath{\sigma}$}})\widehat{\mbox{$\boldmath{\sigma}$}}$ 
are  post-collisional
velocities.
In equation (\ref{2.1}) we have already assumed that, due to the geometry of 
the
problem, the only relevant space coordinate is $z$.
The hydrodynamic fields and the associated fluxes can be expressed as moments
of the distribution function:
\beq
\label{2.2}
n(z,t)=\int d{\bf v}\,  f(z,{\bf v},t),
\eeq
\beq
\label{2.3}
{\sf P}(z,t)=m\int d{\bf v}\,  {\bf v}{\bf v}f(z,{\bf v},t),
\quad p=nk_BT=\frac{1}{3}\mbox{Tr}\,{\sf P},
\eeq
\beq
\label{2.4}
{\bf q}(z,t)=\frac{m}{2}\int d{\bf v}\,  v^2{\bf v} f(z,{\bf v},t).
\eeq
In the expressions of the pressure tensor, equation (\ref{2.3}), and of the
heat flux, equation (\ref{2.4}), we have assumed the absence of a flow 
velocity field. Multiplying equation (\ref{2.1}) by $(v_z,v^2)$ and 
integrating over velocity space, one gets the following {\em steady 
state\/} balance equations
\beq
\label{2.6}
\frac{\partial}{\partial z}P_{zz}+mng=0,
\eeq
\beq
\label{2.7}
\frac{\partial}{\partial 
z}q_{z}=0.
\eeq

In order to solve the Boltzmann equation (\ref{2.1}) one needs to complement 
it with initial and boundary conditions.
Since we will focus on the steady state, the particular choice of initial 
condition is not relevant here. 
As for the boundary conditions, they can be expressed in terms of the 
kernels $K_{\pm}({\bf v}, {\bf v}')$ defined as follows. When a particle 
with velocity ${\bf v}'$ hits the wall at $z=L$, the probability of being 
reemitted with a velocity ${\bf v}$ within the range $d{\bf v}$ is 
$K_{+}({\bf v}, {\bf v}')d{\bf v}$; the kernel $K_{-}({\bf v}, {\bf v}')$ 
represents the same but at $z=0$.
The boundary conditions are then \cite{DB77}
\beqa
\label{2.8}
\Theta(\pm v_z)|v_z| f(z=\{0,L\},{\bf v},t)&=&\Theta(\pm v_z)\int d{\bf v}'\,
|v_z'|K_{\mp}({\bf v}, {\bf v}')\nonumber\\
&&\times\Theta(\mp v_z')f(z=\{0,L\},{\bf v}',t).
\eeqa
In this paper we consider boundary conditions of complete accommodation with 
the walls, so that $K_{\pm}({\bf v}, {\bf v}')=K_{\pm}({\bf v})$ does not 
depend on the incoming velocity ${\bf v}'$ and can be written as
\beq
\label{2.9}
K_{\pm}({\bf v})=A_{\pm}^{-1}\Theta(\mp v_z)|v_z| \phi_{\pm}({\bf v}),\quad
A_{\pm}=\int d{\bf v}\,\Theta(\mp v_z)|v_z| \phi_{\pm}({\bf v}),
\eeq
where $\phi_{\mp}({\bf v})$ represents the probability distribution of a 
fictitious gas in contact with the system at $z=\{0,L\}$. Equation 
(\ref{2.9}) can then be interpreted as meaning that when a particle hits a 
wall, it is absorbed and then replaced by a particle leaving the fictitious 
bath. Of course, any choice of $\phi_{\pm}({\bf v})$ must be consistent with 
the imposed wall temperatures, i.e.
\beq
\label{2.10}
k_BT_{\pm}=\frac{1}{3}m\int d{\bf v}\, v^2 \phi_{\pm}({\bf v}).
\eeq	
The simplest and most common choice is that of a Maxwell-Boltzmann 
distribution:
\beq
\label{2.11}
\phi_{\pm}({\bf v})=\left(\frac{m}{2\pi k_BT_{\pm}}\right)^{3/2}
\exp\left(-\frac{mv^2}{2k_BT_{\pm}}\right).
\eeq
These boundary conditions have been frequently used in molecular dynamics 
simulations \cite{CT80,MKBCN87} as well as in kinetic theory analyses 
\cite{KDSB89,KD89,Cer90,DST99}. Under these conditions, the system is 
understood to be enclosed between two independent baths {\em at 
equilibrium\/} at temperatures $T_+$ and $T_-$, respectively.
While the conditions (\ref{2.11}) are adequate for analysing boundary 
effects \cite{Cer90,W93}, they are not very efficient when the interest lies 
on the transport properties in the bulk.
In order to inhibit the influence of boundary effects, it is more convenient 
to imagine that the two fictitious baths are in {\em nonequilibrium\/} 
states resembling the state of the actual gas near the walls. More 
specifically, we can assume that the fictitious gases are described by the 
BGK equation, whose exact solution for the steady planar Fourier flow (in 
the absence of gravity) is known \cite{SBKD89}. In that case,
\beqa
\label{2.12}
\phi_{\pm}({\bf v})&=&\left(\frac{m}{2\pi k_BT_{\pm}}\right)^{3/2}
\exp\left(-\frac{m(v_x^2+v_y^2)}{2k_BT_{\pm}}\right)
\frac{(2k_BT_{\pm}/m)^{1/2}}{\epsilon_{\pm}|v_z|}\nonumber\\
&&\times
\int_0^\infty d\tau\, \Theta((1-\tau)\mbox{sgn}v_z)\tau^{-3/2}
\nonumber\\
&&\times
\exp\left[-(2k_BT_{\pm}/m)^{1/2}\frac{1-\tau}{\epsilon_{\pm}v_z}-
\frac{mv_z^2}{2k_BT_{\pm}\tau}\right],
\eeqa
where we have additionally assumed statistical independence among the three 
velocity components.
In equation (\ref{2.12}) $\epsilon_{\mp}$ is the local reduced thermal 
gradient at $z=\{0,L\}$. If we formally take the limit $\epsilon_{\pm}\to 
0$, equation (\ref{2.12}) reduces to equation (\ref{2.11}) \cite{SBKD89}. On 
the other hand, if $T_-\neq T_+$, then $\epsilon_{\pm}\neq 0$. The exact 
solution of the  BGK model \cite{SBG86,SBKD89} has the properties $\partial 
T^{3/2}/\partial z=\mbox{const}$ (for hard spheres) and $p=\mbox{const}$; 
from them, it is easy to obtain
\beq
\label{2.13}
\epsilon_{\pm}=\frac{15}{8\sqrt{2\pi}}\frac{T_+^{1/2}-T_-^{1/2}}{\sigma^2L 
\overline{n}}T_{\pm}^{-1/2},
\eeq
where
\beq
\label{2.14}
\overline{n}=\frac{1}{L}\int_0^L 
dz\, n(z)
\eeq
is the average density.
This second class of boundary conditions were first proposed in Ref.\ 
\cite{MASG94}, where they proved to be much more efficient than the 
conditions (\ref{2.11}) for the heat conduction problem. Following the same 
terminology as in Ref.\ \cite{MASG94}, we will refer to the ``equilibrium'' 
conditions (\ref{2.11}) as conditions of Type I and to the 
``nonequilibrium'' conditions (\ref{2.12}) as conditions of Type II.
\section{Simulation method}
\label{sec:3}
In order to solve numerically the Boltzmann equation (\ref{2.1}) with both 
types of boundary conditions, we have used the so-called Direct Simulation 
Monte Carlo (DSMC) method \cite{Bird,Alej}. Comparison with known exact 
solutions of the Boltzmann equation under strong nonequilibrium conditions 
\cite{CC94} proves the reliability and efficiency of the DSMC method to 
solve the Boltzmann equation.
In this method, the velocity 
distribution function is  represented by the velocities $\{{\bf v}_i\}$ 
and positions $\{z_i\}$ of a sufficiently large number of particles $N$. 
Given the geometry of our problem, the physical system is split into layers 
of width $\Delta z$, sufficiently smaller than the mean free path. The 
velocities and coordinates are updated from time $t$ to time $t+\Delta t$, 
where the timestep $\Delta t$ is much smaller 
than the mean free time, by applying a  convection step followed by a 
 collision step.
In the {\em convection step}, the particles are moved 
ballistically, i.e. $v_{iz}\to v_{iz}-g\Delta t$ and $z_i\to 
z_i+v_{iz}\Delta t-\frac{1}{2}g(\Delta t)^2$. In addition, those particles 
crossing the boundaries are reentered with velocities sampled from the 
corresponding probability distribution $K_{\pm}({\bf v})$. 
The {\em 
collision step\/} proceeds as follows \cite{Bird,Alej}. 
For each layer $\alpha$, 
a pair of potential collision partners, $i$ and $j$, and a unit vector
$\widehat{\mbox{$\boldmath{\sigma}$}}_{ij}$ are chosen at random with 
equiprobability.
The collision between particles $i$ and $j$ is then 
accepted with a probability equal to $\Theta({\bf g}_{ij}\cdot
\widehat{\mbox{$\boldmath{\sigma}$}}_{ij})\omega_{ij}/\omega_{\mbox{\scriptsize 
max}}$, where ${\bf g}_{ij}\equiv{\bf v}_i-{\bf v}_j$  is the relative 
velocity, $\omega_{ij}\equiv ({\bf g}_{ij}\cdot
\widehat{\mbox{$\boldmath{\sigma}$}}_{ij})4\pi\sigma^2n_\alpha$ is the 
collision rate of the pair $(i,j)$, $n_\alpha$ being the 
number density in layer $\alpha$, and $\omega_{\mbox{\scriptsize max}}$ is an upper bound 
estimate of the collision rate in the layer. If the collision is accepted, 
post-collisional velocities ${\bf v}_{i,j}'={\bf v}_{i,j}\mp
({\bf g}_{ij}\cdot\widehat{\mbox{$\boldmath{\sigma}$}}_{i})
\widehat{\mbox{$\boldmath{\sigma}$}}_{i}$ are assigned to both particles.
After the collision is processed or if the pair is rejected, the routine 
moves again to the choice of a new pair until the required number of 
candidate pairs $\frac{1}{2}N_\alpha 
\omega_{\mbox{\scriptsize{max}}} \Delta t$ in the layer, where 
$N_\alpha$ is the total number of particles in layer $\alpha$, has been 
processed.

In the course of the simulations, the following ``coarse-grained'' local 
quantities are computed. The number density in layer $\alpha$ is
\beq
\label{2.15}
n_\alpha=\overline{n}\frac{N_\alpha}{(N/L)\Delta z}=\frac{\overline{n} 
L}{N\Delta z}\sum_{i=1}^N
\Theta_\alpha(z_i),
\eeq
where $\Theta_\alpha(z)$ is the characteristic function of layer $\alpha$, 
i.e. $\Theta_\alpha(z)=1$ if $z$ belongs to layer $\alpha$ and is zero 
otherwise.
Similarly, the temperature, the pressure tensor and the heat flux of layer 
$\alpha$ are
\beq
\label{2.16}
k_BT_\alpha=\frac{p_\alpha}{n_\alpha}=\frac{m}{3N_\alpha}\sum_{i=1}^N
\Theta_\alpha(z_i) v_i^2,
\eeq
\beq
\label{2.17}
{\sf P}_\alpha=\frac{L}{N\Delta z}m\sum_{i=1}^N
\Theta_\alpha(z_i) {\bf v}_i{\bf v}_i,
\eeq
\beq
\label{2.18}
{\bf q}_\alpha=\frac{L}{N\Delta z}\frac{m}{2}\sum_{i=1}^N
\Theta_\alpha(z_i) v_i^2{\bf v}_i.
\eeq
According to equation (\ref{2.7}), ${\bf q}$ is a constant in the steady 
state. Thus, we have also evaluated the {\em average\/} heat flux as
\beq
\label{2.19}
\overline{\bf q}=\frac{\Delta z}{L}\sum_\alpha {\bf q}_\alpha=
\frac{1}{N} 
\frac{m}{2}\sum_{i=1}^N
v_i^2{\bf v}_i.
\eeq

The standard definition of mean free path in the case of hard spheres is 
\cite{CC70}
\beq
\label{2.20}
\lambda'=\frac{1}{\sqrt{2} n\pi\sigma^2}.
\eeq	
This quantity is not exactly the same as the mean free path defined by 
equation (\ref{1.5}), which is based on the BGK model. The thermal 
conductivity of a dilute system of hard spheres is (in the first 
approximation) \cite{CC70}
\beq
\label{2.21}
\kappa=\frac{75(\pi mk_BT)^{1/2}k_B}{64 m\pi\sigma^2},
\eeq
so that $\lambda=(15\sqrt{\pi}/16)\lambda'$.
In the following, we take units such that $\overline{\lambda}'\equiv 
(\sqrt{2}\overline{n}\pi\sigma^2)^{-1}=1$ (length unit), $m=1$ (mass unit), 
$(2k_BT_+/m)^{1/2}=1$ (speed unit), $T_+=1$ (temperature unit) and 
$\overline{n}=1$ (density level).
The units of these and other related quantities are given in Table 
\ref{table0}. The table also gives the values of those units taking as a 
reference example a gas with $\sigma=3$~\AA, $m=3\times 10^{-26}$~kg, $T_+= 
 10^{3}$~K and $\overline{n}=4\times 10^{15}$~m$^{-3}$, which are typical of 
 the atmospheric conditions at an altitude of 220~km \cite{C78}.

\begin{table}
\begin{tabular}{lcl}
\hline
Quantity & Unit&Reference value\\
\hline
Temperature ($T$)&$T_+$&$10^{3}$~K\\
Mass ($m$)&$m$&$3\times 10^{-26}$~kg\\
Length ($z$, $L$)&$\overline{\lambda}'$& 625~m\\
Speed ($v$)&$(2k_BT_+/m)^{1/2}$&959~ms$^{-1}$\\
Time ($t$)&$\overline{\lambda}'(2k_BT_+/m)^{-1/2}$& 0.652~s\\
Acceleration ($g$)&$\overline{\lambda}'^{-1}(2k_BT_+/m)$& $1.47\times 
10^{3}$~ms$^{-2}$\\ 
Number density ($n$)&$\overline{n}$&$4\times 10^{15}$~m$^{-3}$\\ 
Pressure ($p$, $P_{zz}$)& $\overline{n}(2k_BT_+)$&$1.10\times 
10^{-4}$~Nm$^{-2}$\\ 
Heat flux ($q_{z}$)& 
$\overline{n}(2k_BT_+)^{3/2}m^{-1/2}$&0.106~Jm$^{-2}$s$^{-1}$\\ 
\hline
\end{tabular}
\caption{
Units used in this paper for the relevant quantities.
The third column gives the ``real'' values of those units taking 
$\sigma=3$~\AA, $m=3\times 10^{-26}$~kg, $T_+= 
 10^{3}$~K and $\overline{n}=4\times~10^{15}$~m$^{-3}$ as a 
reference example. 
\label{table0}}
\end{table}

Each  
physical situation is then characterized by the values of $L$, $T_-$ and $g$ 
only.
In terms of quantities expressed in the above units,
the reduced local parameters $\epsilon$ and $g^*$, defined by equations 
(\ref{1.3}) and (\ref{1.4}), are given as
\beq
\label{2.22}
\epsilon=\frac{5\sqrt{\pi}}{16pT^{1/2}}\left|\frac{\partial 
T^{3/2}}{\partial z}\right|,
\eeq
\beq
\label{2.23}
g^*=\frac{15\sqrt{\pi}}{32}\frac{g}{p}.
\eeq
According to the Fourier law, equation (\ref{1.1}), the heat flux is
\beq
\label{2.24}
q_z^{\text{NS}}=-\frac{25\sqrt{\pi}}{64}\frac{\partial 
T^{3/2}}{\partial z},
\eeq
where we have taken into account that $\kappa\propto T^{1/2}$ for a 
hard-sphere gas and have used the identity $T^{1/2} dT=\frac{2}{3} dT^{3/2}$.
Equation (\ref{2.24}) implies that
in the steady 
state described by the Fourier law one would have  $\partial 
T^{3/2}/\partial z=\mbox{const}$. 
While this  is not in general true, the slope of $T^{3/2}$ is expected to 
change more smoothly than that of $T$. This is why
in equation (\ref{2.22}) $\epsilon$ is written in terms of the former 
rather than in terms of the  latter. 

In the simulations we have taken  $N=5000$ particles, a layer width $\Delta 
z=0.1$ and a timestep $\Delta t=0.004$. We have started from initial 
conditions of the form
\beq
\label{2.25}
f(z,{\bf v},0)=n(z,0)\left(\frac{m}{2\pi k_BT(z,0)}\right)^{3/2}
\exp\left[-\frac{mv^2}{2k_BT(z,0)}\right],
\eeq
where $n(z,0)\propto 1/T(z,0)$ and $T(z,0)=T_-(1+cz/L)^{2/3}$, with $c\equiv 
(T_+/T_-)^{3/2}-1$.
After a time period $t=200$ the system has already relaxed to the steady 
state \cite{MASG94}. We follow the evolution of the system until $t=2000$ 
and average the relevant quantities over 5000 snapshots equally spaced 
between $t=200$ and $t=2000$.
\section{Results}
\label{sec:4}
By using the method outlined in the previous section, we have analysed 
 51 different states. In all of them, 
the separation between the plates has been taken as $L=10$. Two different 
temperature ratios have been considered ($T_-=0.01$ and $T_-=0.05$) and 
boundary conditions of Types I and II have been applied. For each one of 
these four combinations, 12 or 13 different values of $g$ have been 
taken, typically in the range $-0.024<g<0.014$.
We illustrate in figures \ref{fig1}--\ref{fig3bis} the  profiles found in 
the simulations by choosing the case $T_-=0.05$ with boundary conditions of 
Type II as a reference example. The corresponding temperature profiles are 
shown in figure \ref{fig1} for $g=-0.016$, $-0.008$, 0, 0.008 and 0.012. We 
observe that the temperature gradient is larger for $g<0$ (i.e. when the 
gravity field is antiparallel to the heat flux) than for $g>0$ (gravity 
field parallel to the heat flux). This implies that the magnitude of the 
heat flux is expected to be larger in the former case than in the latter. 
This is confirmed by figure \ref{fig2}, where the profile of $q_z$ is 
plotted for the same situations as in figure \ref{fig1}. Figure \ref{fig2} 
also shows that, except for statistical fluctuations, the results are 
consistent with $q_z=\mbox{const}$. This is a consistency test that a steady 
state has been reached in the simulations  [cf.\ equation (\ref{2.7})].
The fact that the heat fluxes are constant and yet the profiles of $T^{3/2}$ 
are nonlinear can be traced back to local deviations from the 
Fourier law.

The profiles of $p$ and $P_{zz}$ are shown in figure \ref{fig3}. At 
$g=0$ the simulation results are consistent with a constant $P_{zz}$.
On the other hand, in agreement with equation (\ref{2.6}), $P_{zz}$ is an 
increasing (decreasing) function of $z$ when $g<0$ ($g>0$). The hydrostatic 
pressure $p$ is slightly larger than $P_{zz}$ for $z/L$ smaller than about 
0.3--0.4, while it is slightly smaller than $P_{zz}$ for larger distances 
from the cold wall. It is worthwhile noting that
the local density $n=2p/T$ is much smaller in the hotter layers 
than in the colder ones, as seen in figure \ref{fig3bis}. Nevertheless,  
this disparity in the population of particles is widely influenced by the 
sign of $g$. Thus, for $g=-0.016$ the densities near the cold and hot walls 
are $n\simeq 4$ and $n\simeq 0.6$, respectively, while those values are 
$n\simeq 8$ and $n\simeq 0.3$ for $g=0.012$. 
The large densities near the cold wall are responsible for the abrupt change 
of the pressure in that region, in agreement with the balance equation 
(\ref{2.6}).

In the cases $T_-=0.05$ (with boundary conditions of Type I) and $T_-=0.01$ 
(with boundary conditions of Types I and II) we have obtained results 
similar to those displayed in figures \ref{fig1}--\ref{fig3bis}. The most 
interesting difference is that, as expected, the boundary effects are much 
less important when applying boundary conditions of Type II than those 
of Type I. This is illustrated in figure \ref{fig4}, where it can be seen 
that, given a value of $L$, $T_-$ and $g$, the jump temperature at the walls 
is much smaller in the case of boundary conditions of Type II. This extends 
to $g\neq 0$ the observation made in Ref.\ \cite{MASG94} for $g=0$.

As stated in section \ref{sec:1}, we are mainly interested in investigating 
the influence of the gravity strength on the heat flux {\em relative\/} to 
the value predicted by the  Navier-Stokes approximation with the actual 
thermal gradient. This ratio is the reduced thermal conductivity $\kappa^*$, 
equation (\ref{1.2}). In the limit in which boundary effects are negligible 
($L\to\infty$), it is expected that $\kappa^*$  depends on position only 
through a functional dependence on the local parameters $\epsilon$ and 
$g^*$, defined by equations (\ref{1.3}) and (\ref{1.4}) or, equivalently, by 
equations (\ref{2.22}) and (\ref{2.23}) in 
our units. In order to minimize as much as 
possible the unavoidable boundary effects associated with a finite $L$, we 
will focus on the region around a point $z_0$  sufficiently far from both 
boundaries. By measuring the pressure, temperature and thermal gradient at 
$z=z_0$ (the heat flux is measured as the average value $\overline{q}_z$), 
we can compute the associated values of $\kappa^*$, $g^*$ and 
$\epsilon=\epsilon_0$.
The question arises as to how to choose the value of $z_0$. 
Here we have applied the following criterion. 
First, the value of $z_0$ 
for $g=0$ is such that the expected number of collisions a particle 
experiences when going from $z=0$ to $z=z_0$ or from $z=L$ to $z=z_0$ is 
5, i.e. ${\cal N}(z_0)=5$, where
\beq
\label{2.26}
{\cal N}(z)=\int_0^z \frac{dz'}{\lambda'(z')}.
\eeq
Notice that, since $\lambda'=1/n$ in our units, ${\cal N}(L)=L=10$.
In this way we have obtained $\epsilon_0=0.33$ ($T_-=0.05$, Type I), 
$\epsilon_0=0.36$ ($T_-=0.05$, Type II), $\epsilon_0=0.44$ ($T_-=0.01$, Type 
I) and $\epsilon_0=0.48$ ($T_-=0.05$, Type II), all of them with $g=0$.
If we followed the same method to choose $z_0$ when $g\neq 0$, then we would 
obtain a different value of $\epsilon_0$ each time and that would hinder the 
analysis about the {\em direct\/} influence of the gravity field on the 
coefficient $\kappa^*$. 
Therefore, as the second part of the criterion, we fix the above values of 
$\epsilon_0$ for each one of the four combinations of $T_-$ and type of 
boundary conditions and then determine $z_0$ to accommodate the 
corresponding $\epsilon_0$. The values of $z_0$, ${\cal N}(z_0)$, $g^*$ and 
$\kappa^*$ obtained in this way are given in Tables 
\ref{table1}--\ref{table4}. As we can see, the point $z_0$ is always closer 
to the cold wall than to the hot wall. However, when the separation is 
measured in terms of the number of collisions rather than as an actual 
distance, it turns out that the point $z_0$ is ``closer'' to the hot wall, 
i.e. ${\cal N}(z_0)>5$, if $g>0$, while the opposite happens if $g<0$.
\begin{table}
\begin{tabular}{cccccc}
\hline
$g$ & $z_0/L$ & ${\cal N}(z_0)$ & $g^*$ & $\kappa^*$& $\Delta\kappa^*$\\
\hline
$-$0.024 & 2.55 & 3.3 & $-$0.095 & 0.903 &$-$0.040\\
$-$0.020 & 2.55 & 3.5 & $-$0.079 & 0.905 &$-$0.038\\
$-$0.016 & 2.60 & 3.8 & $-$0.063 & 0.906 &$-$0.037\\
$-$0.012 & 2.60 & 4.1 & $-$0.048 & 0.917 &$-$0.026\\
$-$0.010 & 2.60 & 4.3 & $-$0.040 & 0.921 &$-$0.022\\
$-$0.008 & 2.55 & 4.3 & $-$0.032 & 0.927 &$-$0.016\\
$-$0.006 & 2.60 & 4.5 & $-$0.025 & 0.934 &$-$0.009\\
$-$0.004 & 2.50 & 4.5 & $-$0.017 & 0.939 &$-$0.004\\
0.000 & 2.70 & 5.0 & 0.000 & 0.943 &0.000\\
0.003 & 2.80 & 5.3 & 0.013 & 0.951 &0.008 \\
0.006 & 2.85 & 5.6 & 0.028 & 0.969 &0.026\\
0.008 & 2.90 & 5.8 & 0.039 & 0.978 &0.035\\
0.010 & 3.20 & 6.3 & 0.052 & 0.986 &0.043\\
\hline
\end{tabular}
\caption{
Values of $z_0$, ${\cal N}(z_0)$, $g^*$, $\kappa^*$ and $\Delta\kappa^*$ for 
different values of $g$ in the case $T_-=0.05$ with boundary conditions of 
Type I. The values of $z_0$ are such that the reduced thermal gradient is 
$\epsilon_0=0.33$.
\label{table1}}
\end{table}
\begin{table}
\begin{tabular}{cccccc}
\hline
$g$ & $z_0/L$ & ${\cal N}(z_0)$ & $g^*$ & $\kappa^*$& $\Delta\kappa^*$\\
\hline
$-$0.020 & 2.45 & 3.9 & $-$0.068 & 0.908 &$-$0.055\\
$-$0.016 & 2.45 & 4.1 & $-$0.055 & 0.921 &$-$0.042\\
$-$0.012 & 2.45 & 4.3 & $-$0.042 & 0.934 &$-$0.029\\
$-$0.010 & 2.40 & 4.4 & $-$0.035 & 0.940 &$-$0.023\\
$-$0.006 & 2.50 & 4.7 & $-$0.022 & 0.949 &$-$0.014\\
$-$0.003 & 2.50 & 4.9 & $-$0.011 & 0.955 &$-$0.008\\
0.000 & 2.50 & 5.0 & 0.000 & 0.963 &0.000\\
0.004 & 2.60 & 5.5 & 0.016 & 0.983 &0.020\\
0.008 & 2.60 & 5.5 & 0.032 & 0.988 &0.025\\
0.010 & 2.90 & 6.2 & 0.046 & 1.016 &0.053\\
0.012 & 3.00 & 6.4 & 0.058 & 1.036 &0.073\\
0.014 & 3.55 & 7.0 & 0.077 & 1.064 &0.101\\
\hline
\end{tabular}
\caption{
Values of $z_0$, ${\cal N}(z_0)$, $g^*$, $\kappa^*$ and $\Delta\kappa^*$ for 
different values of $g$ in the case $T_-=0.05$ with boundary conditions of 
Type II. The values of $z_0$ are such that the reduced thermal gradient is 
$\epsilon_0=0.36$.
\label{table2}}
\end{table}
\begin{table}
\begin{tabular}{cccccc}
\hline
$g$ & $z_0/L$ & ${\cal N}(z_0)$ & $g^*$ & $\kappa^*$& $\Delta\kappa^*$\\
\hline
$-$0.024 & 2.00 & 3.1 & $-$0.108 & 0.833 &$-$0.066\\
$-$0.020 & 2.00 & 3.4 & $-$0.091 & 0.838 &$-$0.061\\
$-$0.016 & 2.00 & 3.6 & $-$0.073 & 0.838 &$-$0.061\\
$-$0.014 & 2.00 & 3.8 & $-$0.065 & 0.849 &$-$0.050\\
$-$0.012 & 1.95 & 3.9 & $-$0.056 & 0.861 &$-$0.038\\
$-$0.008 & 2.00 & 4.3 & $-$0.038 & 0.871 &$-$0.028\\
$-$0.004 & 2.00 & 4.6 & $-$0.020 & 0.880 &$-$0.019\\
0.000 & 2.05 & 5.0 & 0.000 & 0.899 &0.000\\
0.004 & 2.15 & 5.5 & 0.022 & 0.917 &0.018\\
0.006 & 2.40 & 6.0 & 0.036 & 0.921 &0.022\\
0.008 & 2.70 & 6.5 & 0.052 & 0.930 &0.031\\
0.010 & 3.00 & 7.0 & 0.074 & 0.964 &0.065\\
0.011 & 3.50 & 7.5 & 0.092 & 0.994 &0.095\\
\hline
\end{tabular}
\caption{
Values of $z_0$, ${\cal N}(z_0)$, $g^*$, $\kappa^*$ and $\Delta\kappa^*$ for 
different values of $g$ in the case $T_-=0.01$ with boundary conditions of 
Type I. The values of $z_0$ are such that the reduced thermal gradient is 
$\epsilon_0=0.44$.
\label{table3}}
\end{table}
\begin{table}
\begin{tabular}{cccccc}
\hline
$g$ & $z_0/L$ & ${\cal N}(z_0)$ & $g^*$ & $\kappa^*$& $\Delta\kappa^*$\\
\hline
$-$0.024 & 1.85 & 3.4 & $-$0.088 & 0.837 &$-$0.075\\
$-$0.020 & 1.80 & 3.6 & $-$0.075 & 0.851 &$-$0.061\\
$-$0.016 & 1.80 & 3.8 & $-$0.061 & 0.864 &$-$0.048\\
$-$0.012 & 1.80 & 4.1 & $-$0.047 & 0.879 &$-$0.033\\
$-$0.008 & 1.80 & 4.4 & $-$0.032 & 0.888 &$-$0.024\\
$-$0.004 & 1.85 & 4.7 & $-$0.016 & 0.890 &$-$0.022\\
0.000 & 1.90 & 5.0 & 0.000 & 0.912 &0.000\\
0.004 & 1.95 & 5.5 & 0.019 & 0.939 &0.027\\
0.006 & 2.00 & 5.7 & 0.030 & 0.943 &0.031\\
0.008 & 2.10 & 6.0 & 0.043 & 0.974 &0.062\\
0.010 & 2.20 & 6.3 & 0.058 & 0.992 &0.080\\
0.011 & 2.30 & 6.5 & 0.068 & 1.010 &0.098\\
0.012 & 2.70 & 7.1 & 0.084 & 1.047 &0.135\\
\hline
\end{tabular}
\caption{
Values of $z_0$, ${\cal N}(z_0)$, $g^*$, $\kappa^*$ and $\Delta\kappa^*$ for 
different values of $g$ in the case $T_-=0.01$ with boundary conditions of 
Type II. The values of $z_0$ are such that the reduced thermal gradient is 
$\epsilon_0=0.48$.
\label{table4}}
\end{table}

In the absence of gravity ($g=0$), we observe that $\kappa^*$ is smaller 
than 1. This is in part due to a residual influence of boundary effects 
\cite{MASG94}, as indicated by the fact that $\kappa^*$ is closer to 1 with 
boundary conditions of Type II than with those of Type I. 
From the exact solution of the Boltzmann equation for an {\em unbounded\/} 
system of Maxwell molecules \cite{AMN79}, it follows that  $\kappa^*=1$ for 
$g^*=0$ and arbitrary $\epsilon$.
However, it is possible that $\kappa^*(\epsilon,0)$ is slightly smaller than 
1 in the case of hard spheres, even if any boundary effect is removed.
Thus,
in order to characterize 
the {\em pure\/}  gravity-dependence of the effective thermal conductivity 
$\kappa^*$ at a given value of $\epsilon$, we define
\beq
\label{2.27}
\Delta 
\kappa^*(\epsilon,g^*)\equiv\kappa^*(\epsilon,g^*)-\kappa^*(\epsilon,0).
\eeq
The values of $\Delta \kappa^*$ for the corresponding fixed values of 
$\epsilon_0$ are also included in Tables \ref{table1}--\ref{table4} and  
are  plotted in figure \ref{fig5}.
It is observed in  figure \ref{fig5} that, although somewhat 
scattered, the points tend to lie on smooth curves. The behaviour of $\Delta 
\kappa^*$ is in qualitative agreement with the theoretical predictions for 
Maxwell molecules, equations (\ref{1.6})  and (\ref{1.7}). More 
specifically, the departure from the Fourier law, as measured by $\Delta 
\kappa^*$, is positive when the gravity field is parallel to the heat flux 
($g^*>0$), while it is negative when both vectors are mutually antiparallel 
($g^*<0$). In addition, the magnitude of the deviation is larger in the 
former case than in the latter, i.e. $\Delta \kappa^*(\epsilon,g^*)>-\Delta 
\kappa^*(\epsilon,-g^*)$.
From figure \ref{fig5} we can also conclude that the gravity influence is 
less dramatic when the boundary effects are more important, since $|\Delta 
\kappa^*|$ tends to be smaller in the case of boundary conditions of Type I.
From the data corresponding to the boundary conditions of Type II we can 
estimate that $\Delta \kappa^*(\epsilon,g^*)\simeq B \epsilon g^*+\cdots$, 
where the coefficient $B$ has a value between 2 and 2.5. This is about 4 
times smaller than the exact coefficient $B=46/5$ obtained in the case of 
Maxwell molecules. We are not able at present to elucidate which part of 
this discrepancy in the numerical value of $B$ is attributable to boundary 
effects still present in our simulation results and which part is due to the 
role played by the interaction. Besides, when comparing results obtained 
with different interactions, one must have in mind that the choice of 
appropriate dimensionless parameters is not unique. In our case, we have 
defined $\epsilon$, equation (\ref{1.3}), and $g^*$, equation (\ref{1.4}), 
by using the mean free path given by equation (\ref{1.5}), which is based on 
the BGK model. If, on the other hand, we had used the standard mean free 
path of hard spheres, equation   (\ref{2.20}), then we would have $\Delta 
\kappa^*(\epsilon,g^*)\simeq B' \epsilon g^*+\cdots$, where 
$B'=(\lambda/\lambda')^2B\simeq 2.76 B$.

\section{Conclusions}
\label{sec:5}
In this paper we have 
numerically solved the Boltzmann equation
(by means of the DSMC 
method) for a steady heat conduction problem of hard spheres in the presence 
of a gravity field.
The gas is enclosed between two parallel plates separated a distance equal 
to $L=10$ (average) mean free paths.  Two different 
temperature ratios ($T_-/T_+=0.01$ and $T_-/T_+=0.05$) and two alternative 
types of boundary conditions (boundary effects being more important in 
the case of Type I than in the case of Type II) have been 
 considered. For each one of these four possibilities we have applied 12 or 
13 different values of a constant gravity field ${\bf g}=-g \widehat{\bf 
z}$ normal to the plates. The sign criterion is such that $g>0$ means a 
field antiparallel to the thermal gradient (and hence parallel to the heat 
flux vector), while $g<0$ means the opposite.

The first conclusion we draw from the results is that the hydrodynamic 
profiles are rather sensitive to the value of $g$. While the pressure $p$ is 
roughly uniform if $g=0$, it decreases (increases) with $z$ if $g>0$ 
($g<0$), this effect being more important as the magnitude of $g$ grows. 
This is not surprising since it is an extension to nonequilibrium states of 
the equilibrium barometric law $p(z)\propto e^{-mgz/k_BT}$. A less 
obvious effect appears in the case of the temperature profile. If there were 
no temperature jumps at the walls, the temperature of the gas would change 
from $T_-$ at $z=0$ to $T_+$ at $z=L$, irrespective of the value of $g$. 
However, due to unavoidable boundary effects, $T(0)>T_-$ and $T(L)<T_+$.  
Our simulation results show that the temperature jump at the cold (hot) wall 
 decreases as the value of $g$ increases (decreases). For 
 instance, in the case $T_-=0.05$ with 
boundary conditions of Type II,  $T(0)-T_-\simeq0.04$ and $T_+-T(L)\simeq 
0.04$ for $g=-0.016$, while $T(0)-T_-\simeq0.01$ and $T_+-T(L)\simeq 
0.20$ for $g=0.012$. In other words, the temperature jump at a wall 
decreases as the relative density near that wall increases. The 
more positive (negative)  $g$ is, the larger the density is 
near the cold (hot) wall and the smaller the temperature jump is.
Since the temperature jump is more important near the hot wall, a side 
effect of the above discussion is that the (average) thermal gradient across 
the system increases when $g$  decreases, so that it is larger for $g>0$ 
than for $g<0$. A larger thermal gradient implies a larger magnitude of 
the heat flux and this expectation is confirmed by our simulation results, 
which show that $|q_z|$ clearly increases as $g$ decreases. 

Our interest has not focused, however, on the {\em absolute\/} change of the 
heat flux due to the gravity field, but on its change {\em relative\/} to 
the Navier-Stokes prediction when the {\em actual\/} temperature gradient is 
considered. To that end we have introduced the ratio $\kappa^*$ defined by 
equation (\ref{1.2}), which is a local quantity that in the {\em bulk\/} 
region is expected to depend on space only through a functional dependence 
on the reduced thermal gradient $\epsilon$, equation (\ref{1.3}), and 
gravity strength $g^*$, equation (\ref{1.4}). For each one of the four 
different combinations of $T_-/T_+$ and boundary conditions we have fixed a 
value  $\epsilon=\epsilon_0$ ($\epsilon_0\simeq 0.3$--$0.5$) and have 
analysed the $g^*$-dependence of $\Delta\kappa^*(\epsilon_0,g^*)\equiv 
\kappa^*(\epsilon_0,g^*)-\kappa^*(\epsilon_0,0)$.
The simulation results for hard spheres presented in this paper are in 
qualitative agreement with those obtained for Maxwell molecules by a 
perturbation analysis of the Boltzmann  \cite{TGS97} and the BGK 
 \cite{TGS99b} equations. More specifically, when the field and the heat 
 flux 
are parallel ($g^*>0$) the gravity induces an enhancement of the (relative) 
heat conduction ($\Delta\kappa^*>0$), while the opposite happens when both 
vectors are mutually antiparallel. In addition, the influence of gravity is more 
pronounced in the former case than in the latter. At a quantitative level, 
on the other hand, the values of $|\Delta\kappa^*|$ reported here 
 are typically smaller than those theoretically estimated for comparable 
 values of $\epsilon$ and $g^*$.
A certain part of this difference is possibly due to boundary effects still 
present in our simulations and absent in the theoretical analyses of Refs.\ 
\cite{TGS97} and \cite{TGS99b}. This expectation is supported by the fact 
that $|\Delta\kappa^*|$ is generally smaller in 
the case of conditions of Type I than in that of Type II, thus indicating
that boundary effects tend to mitigate the influence of gravity.
Notwithstanding this, the remaining difference leads us to conclude that  
the response of the system to the application of the field, as measured by  
$\Delta\kappa^*$, is less important in the case of hard spheres than in the 
case of Maxwell molecules.

The authors are grateful to J. M. Montanero for his help in the simulation 
method.
This work has been done under the auspices of the Agencia Espa\~nola de 
Cooperaci\'on Internacional (Programa de Cooperaci\'on Interuniversitaria 
Hispano-Marroqu\'{\i}).
A.S. acknowledges partial support from the DGES (Spain) through 
Grant No.\ PB97-1501 and from the Junta de Extremadura (Fondo Social Europeo)
through Grant No.\ IPR98C019.

\newpage
\begin{figure}
\caption{Schematic illustration of the system geometry.
\label{fig0}}
\end{figure}

\begin{figure}
\caption{Profile of $T^{3/2}$  in the case $T_-=0.05$ with 
boundary conditions of Type II. The values of $g$ are, from 
top to bottom, $g=-0.016$, $-0.008$, 0, 0.008 and 0.012.
The labels C and H denote the locations of the cold and hot walls, 
respectively.
\label{fig1}}
\end{figure}
\begin{figure}
\caption{Profile of $q_z$  in the case $T_-=0.05$ with 
boundary conditions of Type II. The values of $g$ are, from 
bottom to top, $g=-0.016$, $-0.008$, 0, 0.008 and 0.012.
The labels C and H denote the locations of the cold and hot walls, 
respectively.
\label{fig2}}
\end{figure}
\begin{figure}
\caption{Profiles of $p$ (solid lines) and $P_{zz}$  (dashed lines) in the 
case $T_-=0.05$ with boundary conditions of Type II. The values of $g$ are, 
from  top to bottom at the right end, $g=-0.016$, $-0.008$, 0, 0.008 and 
0.012.
The labels C and H denote the locations of the cold and hot walls, 
respectively.
\label{fig3}}
\end{figure}
\begin{figure}
\caption{Profile of $n$  in the 
case $T_-=0.05$ with boundary conditions of Type II. The values of $g$ are, 
from  top to bottom at the right end, $g=-0.016$, $-0.008$, 0, 0.008 and 
0.012. Note that the vertical axis is in logarithmic scale.
The labels C and H denote the locations of the cold and hot walls, 
respectively.
\label{fig3bis}}
\end{figure}
\begin{figure}
\caption{Profile of $T^{3/2}$  in the case $T_-=0.01$ with 
boundary conditions of Types I (dashed lines) and II (solid lines). The 
values of $g$ are (a) $g=-0.016$ and (b) $g=0.012$.
The labels C and H denote the locations of the cold and hot walls, 
respectively.
\label{fig4}}
\end{figure}
\begin{figure}
\caption{Plot of $\Delta \kappa^*$ versus the reduced field strength $g^*$ 
for $T_-=0.05$ with boundary conditions of Type I (open squares, 
$\epsilon=0.33$),
$T_-=0.05$ with boundary conditions of Type II (solid squares, 
$\epsilon=0.36$),
$T_-=0.01$ with boundary conditions of Type I (open circles, 
$\epsilon=0.44$) and
$T_-=0.01$ with boundary conditions of Type II (solid circles, 
$\epsilon=0.48$). The lines are polynomial fits to guide the eye.
\label{fig5}}
\end{figure}

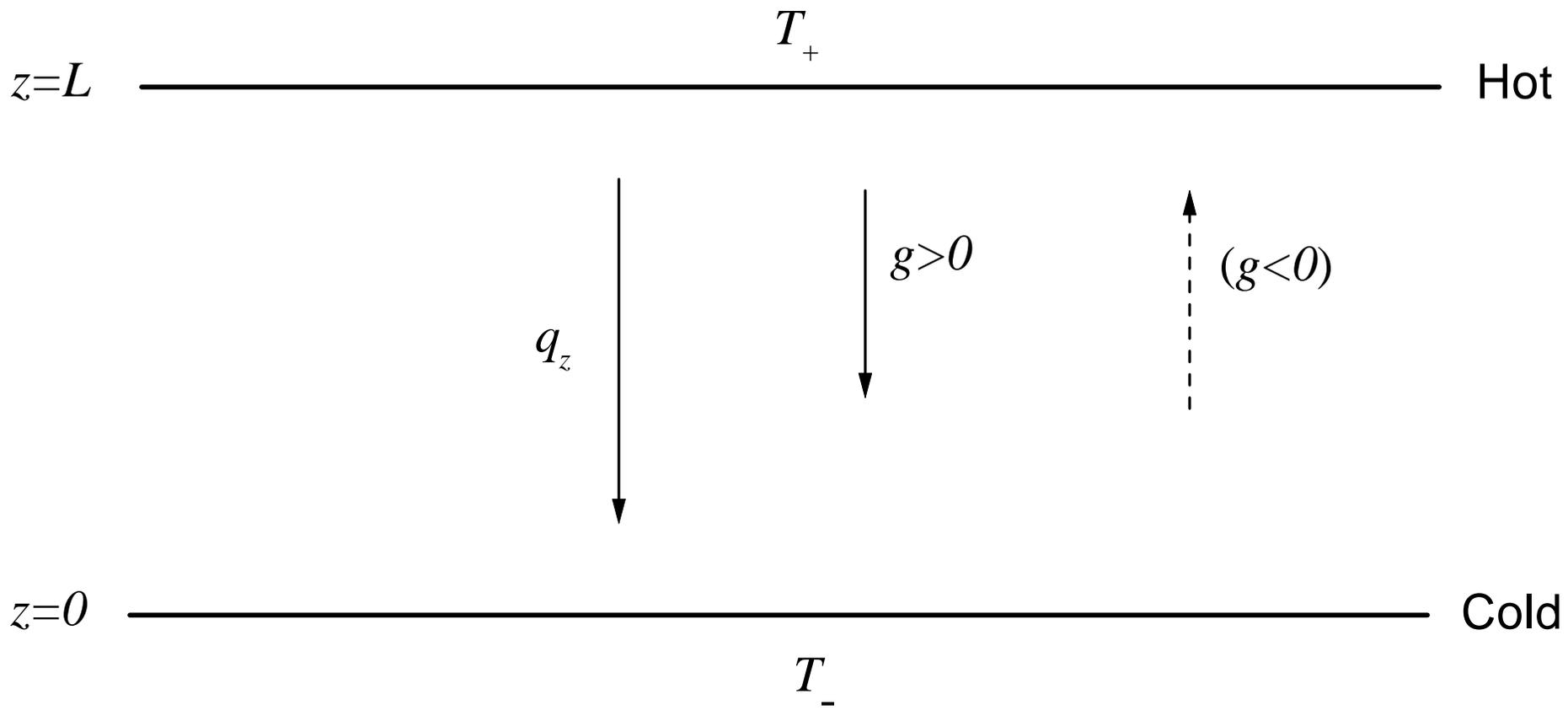
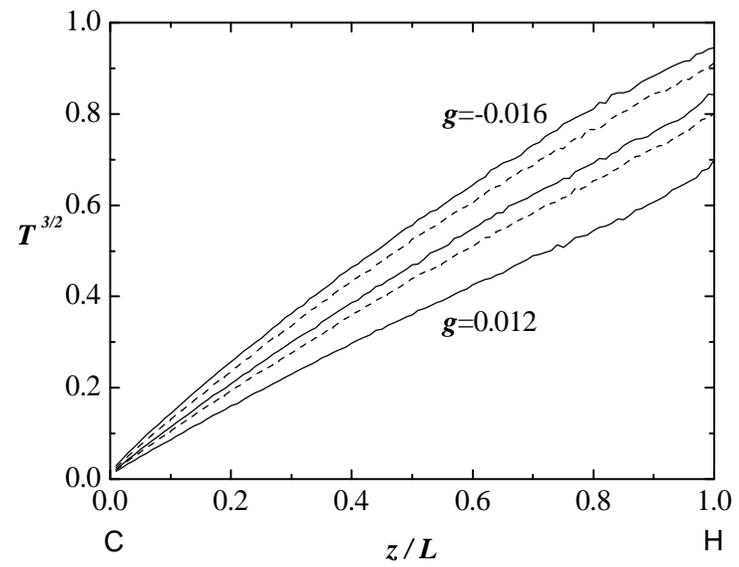
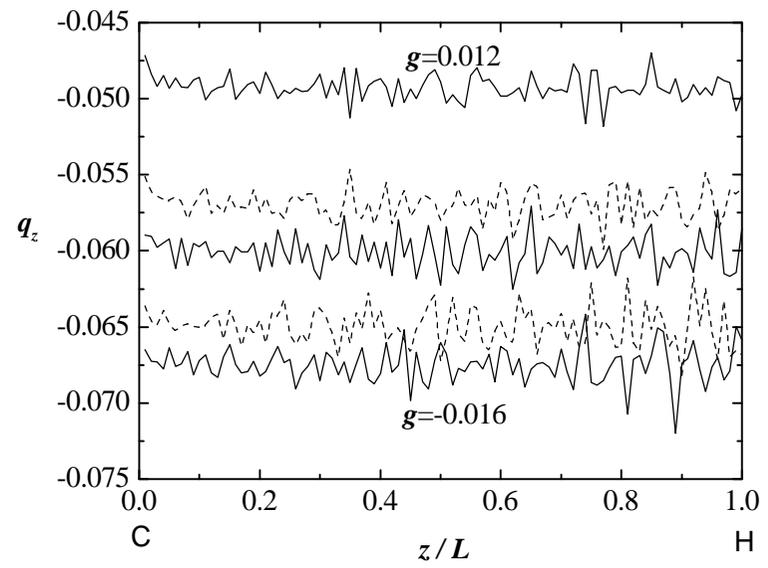
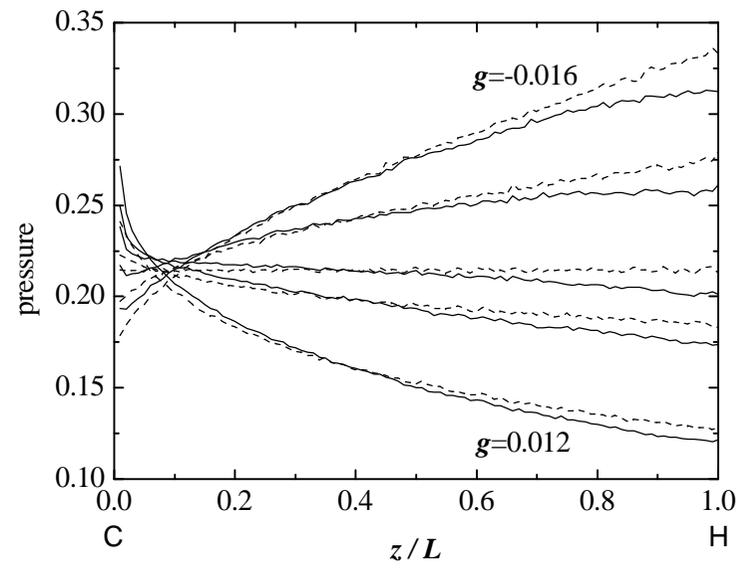
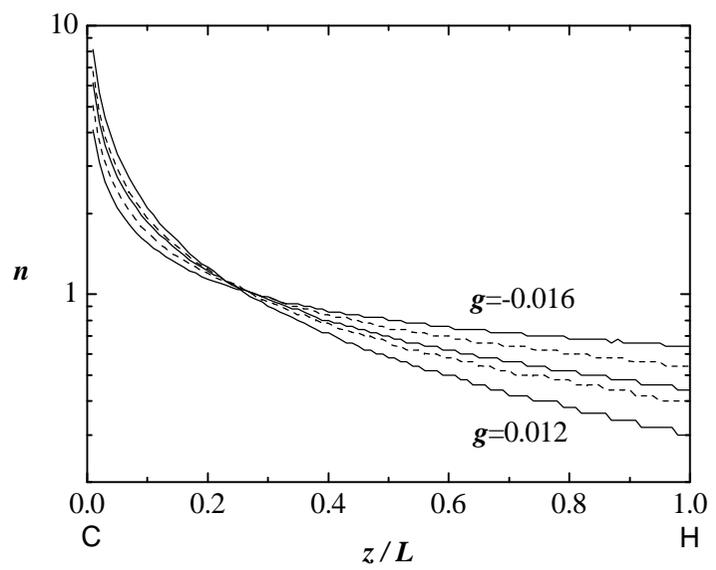
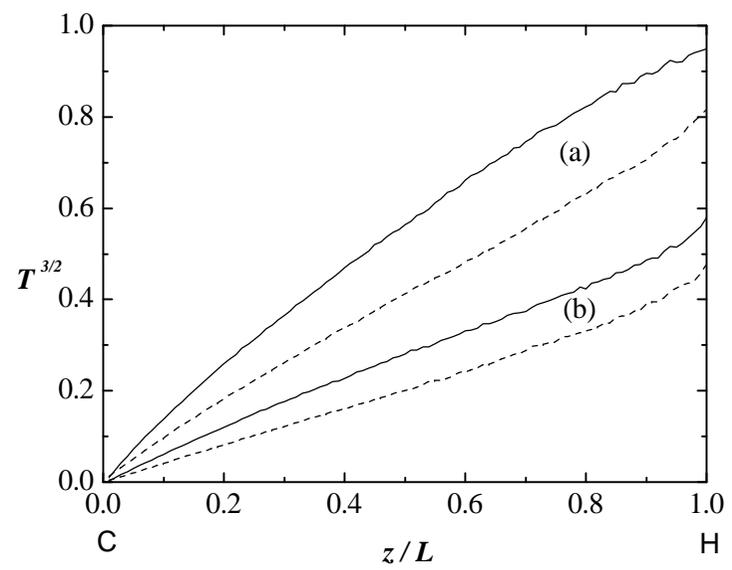
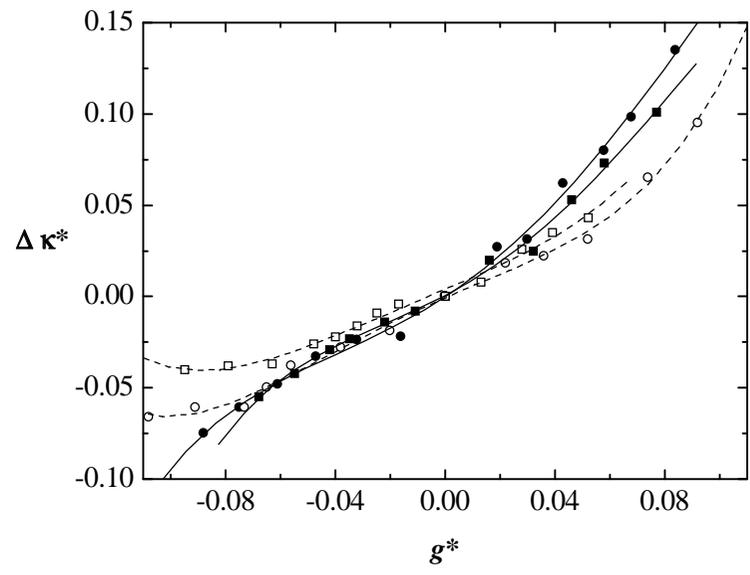
\end{document}